\documentstyle[aps,prl
,preprint
]{revtex}
\begin{document}
\draft
\title{Re-entrant ferroelectricity in liquid crystals}
\author{D. Pociecha$^{1}$, E. Gorecka$^{1}$, M. \v Cepi\v c$^{2}$, N. Vaupoti\v c$%
^{3}$, B. \v Zek\v s$^{2}$, D. Kardas$^{1}$ and J. Mieczkowski$^{1}$}
\address{$^{1}$Chemistry Department, Warsaw University,\\
Al. Zwirki i Wigury 101, 02-089 Warsaw, Poland.\\
$^{2}$J. Stefan\\
Institute, Jamova 39, 1000 Ljubljana, Slovenia;\\
$^{3}$Faculty of\\
Education, Koro\v ska 160, 2000 Maribor, Slovenia;}
\date{\today}
\maketitle

\begin{abstract}
%\widetext
The ferroelectric (Sm~C$^*$) -- antiferroelectric (Sm~C$^*_A$) -- reentrant
ferroelectric (re~Sm~C$^*$) phase temperature sequence was observed for
system with competing synclinic - anticlinic interactions. The basic
properties of this system are as follows (1) the Sm~C$^*$~ phase is
metastable in temperature range of the Sm~C$^*_A$~ stability (2) the double
inversions of the helix handedness at Sm~C$^*$~ -- Sm~C$^*_A$~ and Sm~C$^*_A$%
~ -- re-Sm~C$^*$~ phase transitions were found (3) the threshold electric
field that is necessary to induce synclinic ordering in the Sm~C$^*_A$ phase
decreases near both Sm~C$^*_A$~ -- Sm~C$^*$~ and Sm~C$^*_A$~ -- re-Sm~C$^*$~
phase boundaries, and it has maximum in the middle of the Sm~C$^*_A$~
stability region. All these properties are properly described by simple
Landau model that accounts for nearest neighboring layer steric interactions
and quadrupolar ordering only.
\end{abstract}


%\twocolumn[\hsize\textwidth\columnwidth\hsize\csname@twocolumnfalse\endcsname

%]
%\narrowtext

There are two types of tilted smectic phases with liquid like in-layer
order, synclinic (Sm~C) phase in which molecules are tilted in all layers in
the same direction, and anticlinic (Sm~C$_A$) phase in which tilt direction
alternates when going from one smectic layer to another. For the chiral
tilted systems the spontaneous electric polarization in the smectic layers
along the ${\bbox{c}}\times\bbox{n}$~ direction is allowed, where $\bbox{c}$%
~ is projection of the tilt onto the smectic layer and $\bbox{n}$ is the
layer normal. In result the synclinic Sm~C phase acquires ferroelectric (Sm~C%
$^*$) properties. The anticlinic SmC$_A$ phase becomes antiferroelectric
phase (Sm~C$^*_A$) since the $P_s$ is canceled out in two consecutive layers
as the $\bbox{c}$~ vector direction alternates\cite{ref1}. For some liquid
crystals the free energy of the synclinic and the anticlinic structure is
very similar and the type of the tilted phase formed can be tuned by very
slight modifications in molecular structure. It could be observed that
within the same homologue series, when the length of the terminal alkyl
chain is changed, some homologues exhibit synclinic (ferroelectric) Sm~C$^*$%
~ while the others anticlinic (antiferroelectric) Sm~C$^*_A$~ phase, and
usually more complicated behavior than simple odd-even function of alkyl
chain length takes place\cite{ref2}. When competing interactions of
comparable strength are present, one can also expect more complex
temperature sequence of polar phases. In chiral materials, between synclinic
and anticlinic phases the number of intermediate phases with ferrielectric
order are sometimes detected\cite{ref3}. The competing synclinic --
anticlinic interactions might also lead to the re-entrancy of the
ferroelectric order. We will show that when the temperature is lowered,
instead of simple Sm~C$^*$~ -- Sm~C$^*_A$~ phase sequence, the Sm~C$^*$~
phase re-appears below Sm~C$^*_A$~ phase.

On the phase diagram of the {\it mPIRn}\cite{ref4}, with chemical formula
given on Fig.~\ref{fig5}, $m$~ is number of carbon atoms in achiral tail and
$n$~ is number of carbon atoms in a chain attached to the chiral center
homologue series, in which $m$~ is fixed to 8 and $n$~ is changed (Fig.\ref
{fig1}), the phase sequence Sm~C$^*$~ -- Sm~C$^*_A$~ -- re-Sm~C$^*$~--Sm~I
is observed for compounds with $n=2,3$~ and Sm~C$^*$~ -- Sm~C$^*_A$~ --
re-Sm~C$^*$~ -- Sm~B$_{cry}$~ for $n=4$. In materials with longer chiral
tail, $n=5,6$ the Sm~C$^*_A$~ phase disappears and only one tilted phase,
ferroelectric Sm~C$^*$~ is present. When the chiral end is fixed to $n=4$~
and the non-chiral end is changed (inset Fig.\ref{fig1}) we observed
increase of the Sm~C$^*_A$~ phase temperature range with increasing $m$;
homologue $m=9$~ exhibits the sequence Sm~C$^*$~ -- Sm~C$^*_A$~ -- Sm~I$_{A}$%
~ -- Sm~I, which means that the ferroelectric properties are restored in the
hexatic phase. The phase diagram for {\it mPIRn} series were constructed
based on measurements performed at slow temperature scans (up to 5 K min$%
^{-1}$). However, it should be noticed that under sufficiently fast cooling
or heating, sometimes the antiferroelectric phase was missing, and only
ferroelectric phases were observed. For chosen homologues, it has been
checked that enantiomeric and racemic samples have the same sequence of the
synclinic Sm~C and the anticlinic Sm~C$_A$ phases and that the phase
transition temperatures coincide. More detailed studies have been performed
on the homologues which do not exhibit the reentrant ferroelectric
properties in the hexatic phase.

The enthalpy changes detected in calorimetric measurements (Perkin Elmer DSC
7) at Sm~C$^*$~ -- Sm~C$^*_A$~ and Sm~C$^*_A$~ -- re-Sm~C$^*$~ (eg. for {\it %
8PIR4}) phase transitions were comparable, which suggests that the
reconstructing of the azimuthal angle structure does not involve the changes
of the other order parameters.

The ferroelectric and antiferroelectric properties of the phase could be
distinguished by electric switching, dielectric and optical methods. In the
Sm~C$^*$~ phase bistable switching is detected, while in the Sm~C$^*_A$~
phase tristable switching with a threshold field is observed. The threshold
field in Sm~C$^*_A$~ phase critically decreases near both Sm~C$^*$~ -- Sm~C$%
^*_A$~ and Sm~C$^*_A$~ -- re-Sm~C$^*$~ phase transitions, which shows that
the strength of the anticlinic interactions decreases. The type of
electrooptic switching is consistent with the detected current signal, the
single or double current peak was found upon reversing the spontaneous
polarization ($P_{s}$) by electric field of low frequency ($\sim 1$ Hz), in
Sm~C$^*$~ and Sm~C$^*_A$~ phases, respectively. Moreover, both Sm~C$^*$~ --
Sm~C$^*_A$~ and Sm~C$^*_A$~ -- re-Sm~C$^*$~ phase transitions are marked by
notable changes in the texture, related to the changes of the birefringence
and the helical structure. Except for shortest homologue, in {\it 8PIRn}
series in synclinic phases the helical structure gives selective reflection
in visible light range (eg. $\sim 610$~nm for 8PIR4), while in anticlinic
phase the helical pitch is of few microns length. For the shortest homologue
{\it 8PIR2} in both Sm~C$^*$~and re-Sm~C$^*$~ phases as well as in the Sm~C$%
^*_A$~ phase the pitch is longer than visible light wavelength, since
significant optical activity is detected. The optical activity changes sign
when going from the Sm~C$^*$~ phase to the Sm~C$^*_A$~ phase, having the
same sign in both Sm~C$^*$~ and re-Sm~C$^*$~ phases. This indicates that the
sense of the helical structure alternates when going from synclinic to
anticlinic phase.

No tilt anomaly was detected at Sm~C$^*$~ -- Sm~C$^*_A$~ -- re-Sm~C$^*$~
phase transitions. The tilt angle sligthly increases on cooling over all
temperature range of tilted phases (eg. from 24$^o$ to 28$^o$ and from 22$^o$
to 25$^o$ for {\it 8PIR2} and {\it 8PIR4}, respectively), without any
pretransitional anomaly neither at phase transition to hexatic smectic I nor
to crystalline smectic B phase. Also the $P_s$ value does not exhibit
anomaly at the Sm~C$^*$~ -- Sm~C$^*_A$~ phase transitions.

Dielectric measurements were carried out with Wayne Kerr impedance analyzer
for glass cells of various thickness (2 - 100 microns) covered with
indium-tin-oxide (ITO) transparent electrodes and parallel rubbed polyimide
layer. In thick, 25 and 100 micron cells the Goldstone (phason) mode with
the relaxation frequency in kHz regime was detected in both ferroelectric
Sm~C$^*$~ and re-Sm~C$^*$~ phases (Fig.\ref{fig2}a). The transition to
antiferroelectric Sm~C$^*_A$~ phase is marked by sudden suppression of the
phason mode, where no dielectric mode was found within the studied frequency
range (20 Hz - 300 kHz).

In thin, 2 micron, cells in which interactions from the surfaces favoring
ferroelectric order become more significant, we observed suppression of the
antiferroelctric order (Fig.\ref{fig2}b). In such cells in the whole range
of tilted smectic phases the Goldstone mode was detected, with significantly
higher relaxation frequency than observed in thicker cells\cite{ref5}. In
cells with intermediate thickness, in the temperature range of SmC$_A$ phase
stability, the clusters of both Sm~C$^*$~ and Sm~C$^*_A$~ phases are
observed, and the number of Sm~C$^*$~ clusters increases when the thickness
decreases\cite{ref6}.

The proper order parameter to describe transition between tilted smectic
phases is vector given by $\bbox{\xi}_{j}=\theta \,\left( \cos \varphi
_{j},\sin \varphi _{j}\right) $, where $\theta $ is the magnitude of the
tilt and the azimuthal angle $\varphi _{j}$ describes the tilt direction in
the $j$-th layer. Since the synclinic -- anticlinic -- synclinic phase
sequence is observed in enantiomers and their racemate mixture, we can
assume that for the studied system the chiral interactions as well as polar
interactions between layers are negligible. Thus the Landau free energy
density expansion can be written as
\begin{eqnarray}
G &=&\sum \left( \frac{1}{2}\;a_{0}\;\bbox{\xi}_{j}^{2}+\right. \frac{1}{4}%
\;b_{0}\;\bbox{\xi}_{j}^{4}+\frac{1}{6}\;c_{0}\;\bbox{\xi}_{j}^{6}+
\nonumber \\
&&\frac{1}{2}\;a_{1}\;\left( \bbox{\xi}_{j}\cdot \bbox{\xi}_{j+1}\right) +%
\frac{1}{2}\;a_{1}^{\prime }\;\bbox{\xi}_{j}^{2}\;\left( \bbox{\xi}_{j}\cdot %
\bbox{\xi}_{j+1}\right) +  \nonumber \\
&&\left. \frac{1}{2}\;a_{1}^{\prime \prime }\;\bbox{\xi}_{j}^{4}\;\left( %
\bbox{\xi}_{j}\cdot \bbox{\xi}_{j+1}\right) +\frac{1}{4}\;b_{1}\;\left( %
\bbox{\xi}_{j}\cdot \bbox{\xi}_{j+1}\right) ^{2}\right)
\end{eqnarray}
where the only temperature dependent parameter is $a_{0}=a(T-T_{0})$, $T_{0}$
being transition temperature from orthogonal to the tilted smectic phase in
the absence of interlayer correlations. The first three terms with the
parameters $a_{0},b_{0}$ and $c_{0}$ model intralayer interactions and
resume the fact, that the transition to the tilted phase is of the strong
first order and that the tilt over the whole stability region of the Sm~C$%
^{*}$~ and the Sm~C$_{A}^{*}$~ phase does not change significantly.
Interlayer interactions are given by the terms $a_{1},a_{1}^{\prime
},a_{1}^{\prime \prime }$. Quadrupolar interlayer interactions are given by
the $b_{1}$ term. Introducing new parameter $\alpha _{j}=\alpha =\varphi
_{j+1}-\varphi _{j}$ which is 0 for the Sm~C$^{*}$~ phase and $\pi $ for Sm~C%
$_{A}^{*}$~ phase, the free energy (1) can be re-written as
\begin{eqnarray}
G &=&\frac{1}{2}\;a_{0}\;\theta ^{2}+\frac{1}{4}\;b_{0}\;\theta ^{4}+\frac{1%
}{6}\;c_{0}\;\theta ^{6}+  \nonumber \\
&&\frac{1}{2}\;\left( a_{1}+a_{1}^{\prime }\;\theta ^{2}+a_{1}^{\prime
\prime }\;\theta ^{4}\right) \theta ^{2}\;\cos \;\alpha +  \nonumber \\
&&\frac{1}{4}\;b_{1}\;\theta ^{4}\;\cos ^{2}\alpha .
\end{eqnarray}
The effective interlayer interactions are determined by sign of $%
a_{1ef}=(a_{1}+a_{1}^{\prime }\;\theta ^{2}+a_{1}^{\prime \prime }\;\theta
^{4})$, if negative the interlayer interactions enforce synclinic tilt in
the neighboring layers, if positive anticlinic structure is favored. On
microscopic level, the sign of $a_{1ef}$ term results from competing
interactions e.g. molecular interpenetration through the layers benefits
negative sign while strong attractive van der Waals interactions between
molecules from neighboring layers benefit positive sign. Positive sign also
favour antiparallel dipolar ordering of in neighboring layers and can be
significant in systems with strong electrostatic interactions \cite{ref7}.

The observed re-appearance of the Sm~C$^{*}$ phase raises the question what
is the driving force to this phenomenon. From the expression for $a_{1ef}$
can be deduced that reentrant phenomenon could be observed in systems where $%
a_{1ef}$ changes its sign at two different temperatures. This could happen
in systems where either the tilt $\theta $~or the spontaneous polarization $%
P_s$ (or both) depend on the temperature nonmonotonically. However, we have
to exclude both factors. In studied system tilt steadily grows with lowering
the temperature and the synclinic -- anticlinic -- re-synclinic phase
sequence is also observed in racemic compounds forming non-polar mesophases.

Thus the other interactions have to be explored as a driving mechanism for
the re-entrancy of the Sm~C$^{*}$~ phase. At higher temperatures one can
expect negative sign of $a_{1ef}$ due to the strong diffusion of molecules
between layers. At lower temperatures the quadrupolar ordering can change
interactions between molecules from neighboring layer. If the molecules are
considered as flat, lath-like species, their quadrupolar ordering means
correlations between direction of short molecular axes. For the tilted phase
as the biaxial ordering becomes more important with decreasing temperature,
the molecules arrange with their short axes perpendicular to the tilt
direction \cite{ref8}. This type of ordering might promote the interlayer
diffusion of molecules and consequently the re-appearance of the Sm~C$^{*}$~
phase.

It should be stressed that although nonmotonic dependence of $a_{1ef}$ can
also be observed in other tilted phases \cite{ref9}, but only in the systems
where the free energy part of synclinic and anticlinic interactions are
similar in a broad temperature range, the free energies $G_{\alpha =0}(T)$
and $G_{\alpha =\pi }(T)$ can become equal at two different temperatures,
thus re-entrant phenomenon, could be observed.

When molecules order quadrupolarly as discussed, also the sign of $b_{1}$
term defining quadrupolar interlayer interactions is negative, since tilts
bound to one plane are preferred and therefore the free energy has always
two minima as a function of $\alpha $. The behavior has been observed also
in some other materials \cite{ref10}. Appearance of the local minimum
explains why in temperature range when the anticlinic structure is
thermodynamically stable, the metastable synclinic phase can be sometimes
observed.

In order to analyze hellicoidal modulations, in chiral system the Lifshitz
term which account for chiral interlayer interactions should be added to the
free energy.
\begin{equation}
G_{chir}=G+f \;\left( \bbox{\xi}_{j}\times \bbox{\xi}_{j+1}\right)_z =G+f
\;\theta ^{2}\;\sin \alpha.
\end{equation}
For weak chiral interactions we can assume that $\alpha \approx 0$ in Sm~C$%
^* $~ phase, and $\alpha \approx \pi $ in the Sm~C$^*_A$~ phase. Minimizing
the free energy with respect to $\alpha $ one obtains $\alpha =\frac f{%
a_{1ef}+b_1\theta ^2}$~ for the Sm~C$^*$~ phase and $\alpha =\pi -\frac f{%
-a_{1ef}+b_1\theta ^2}$~ for the Sm~C$^*_A$~ phase, respectively. The sign
of $f$ correlates with enantiomer chirality. If in the Sm~C$^*$~ phase $%
\alpha> 0$ then it is slightly less than $\pi $ (since $a_{1ef}$ reverses
sign when going from Sm~C$^*$~ to Sm~C$^*_A$) in the Sm~C$^*_A$~ phase, thus
opposite handedness of the helices in the Sm~C$^*$~ phase and the Sm~C$^*_A$%
~ phase is expected as experimentally observed.

To account for the system interaction with electric field, the term $-%
\bbox{E}\cdot \bbox{P}_j$ is added to the free energy $G$, where $\bbox{P}_j$
is the polarization in the $j$-th layer. The straightforward calculations
show that the energies of the untwisted Sm~C$^{*}$~ and Sm~C$_A^{*}$~ phases
become equal when the electric field induced switching is observed at $%
E_{th} $~ given by $E_{th}\;P_s=a_{1ef}\;\theta ^2$, where $P_s$ is the
spontaneous polarization. In the Sm~C$_A^{*}$~ phase the threshold field $%
E_{th}$ therefore follows the $a_{1ef}$ temperature dependence (Fig.\ref
{fig3}), it is largest in the middle of the temperature range of the Sm~C$%
_A^{*}$~ phase stability and decreases near both Sm~C$_A^{*}$~ -- Sm~C$^{*}$%
~ phase transitions. Fitting simultaneously tilt $\theta $ vs. temperature
and $a_{1ef}=\frac{E_{th}\;P_s}{\theta ^2}$~ vs. $\theta $ provided for
compound {\it 8PIR2} the following set of the parameters describing the
strength of molecular interactions, in-layer and nearest neighboring
interlayer $a_0=1.2$\ $kJ\ mol^{-1}K^{-1},b_0+b_1=-146.0\ kJ\
mol^{-1},c_0=1003.5\ kJ\ mol^{-1},a_1=-0.16\ kJ\ mol^{-1},a_1^{\prime
}=1.51\ kJ\ mol^{-1},a_1^{\prime \prime }=-3.52$\ $kJ\ mol^{-1}$.The
temperature dependence of the tilt $\theta =\sqrt{\frac{-B+\sqrt{B^2-4AC}}{2C%
}}$with $A=a(T-T_0)-a_1,B=b_0+b_1-2a_1^{\prime }$ and $C=c_0-3a_1^{\prime
\prime }$ was obtained by minimizing free energy (2) in anticlinic phase
with respect to the tilt. The parameters $a_0=0.2$\ $kJ\
mol^{-1}K^{-1},b_0+b_1=-11.7\ kJ\ mol^{-1},c_0=343.3\ kJ\
mol^{-1},a_1=-0.02\ kJ\ mol^{-1},a_1^{\prime }=0.32\ kJ\
mol^{-1},a_1^{\prime \prime }=-0.73$\ $kJ\ mol^{-1}$ were obtained if the
similar procedure is applied in the Sm~C$_A^{*}$ phase of the prototype
antiferroelectric material {\it MHPOBC}\cite{ref1} (Fig.\ref{fig3}). For
above parameters, it has been estimated that the interlayer interactions
contribution to the free energy is less than 0.03\% for {\it 8PIR2 }and
0.3\% for {\it MHPOBC }compound, which clearly indicates that the interlayer
interactions in compounds {\it mPIRn} series are much weaker than interlayer
inetractions in materials without reentrant behavior.

Summarizing, the system with temperature sequence of synclinic (Sm~C$^*$) --
anticlinic (Sm~C$^*_A$) -- synclinic (Sm~C$^*$) tilted smectic phases was
found. It has been proved that the unusual phase sequence is related to
steric interactions since it is observed in enetiomeric as well as in
racemic compounds. The Sm~C$^*_A$~ phase can be easily suppressed by surface
interactions in thin cells. The Sm~C$^*$~ state could be observed as a
metastable phase in temperature range Sm~C$^*_A$~ thermodynamic stability.
The simple Landau model that accounts only interactions with nearest
neighboring layesr and quadrupolar ordering is able to explain the Sm~C$^*$~
-- Sm~C$^*_A$~ -- re-Sm~C$^*$~ phase sequence and properties of {\it mPIRn}
compounds. It resumes observed in experiment fact that the helix in Sm~C$^*$%
~ and re-Sm~C$^*$~ phases has opposite sign than in Sm~C$^*_A$~ phase. It
also explains that the threshold field which induce synclinic ordering in
the Sm~C$^*$~phase decreases near both Sm~C$^*_A$~ -- Sm~C$^*$~ and Sm~C$%
^*_A $~ -- re-Sm~C$^*$~ phase boundaries, and has maximum in the middle of
the Sm~C$^*_A$~ region.

\medskip

The work was supported by KBN Grant No. 3T09A 046 15. The financial support
of the Slovenian Ministry of science is acknowledged.

\begin{figure}[tbp]
\caption{Chemical formula of the material {\it mPIRn }.}
\label{fig5}
\end{figure}

\begin{figure}[tbp]
\caption{Phase diagram for homologue series {\it 8PIRn }in which the length
of alkyl chain attached to the chiral carbon atom is changes, in the inset
phase diagram for {\it mPIR4 }in which the length of the achiral alkoxy
chain is changed.}
\label{fig1}
\end{figure}

\begin{figure}[tbp]
\caption{3-D temperature-frequency plot of the real part of dielectric
constant $\varepsilon ^{*}$ for compound {\it 8PIR4 }measured in 100$\mu m$
(a) and 2$\mu m$ (b) thick cells. }
\label{fig2}
\end{figure}

\begin{figure}[tbp]
\caption{Tilt dependence of $a_{1ef}$ effective interlayer interactions for
{\it 8PIR2 }and {\it MHPOBC, } in the Sm~C$^*_A$~ phase. At the transition
Sm~C$^*$~ --Sm~C$^*_A$~ and Sm~C$^*_A$~ -- re-Sm~C$^*$~, the $a_{1ef}$
changes sign, thus stabilizes the synclinic Sm~C$^*$~ phase.}
\label{fig3}
\end{figure}

\end{document}